\documentclass[11pt,onecolumn,oneside,peerreview]{IEEEtran}

\IEEEoverridecommandlockouts
\usepackage{cite}
\usepackage{graphicx}
\usepackage{amsmath}
\usepackage{times}
\usepackage{latexsym}
\usepackage{bm}
\usepackage{amssymb}
\usepackage[center]{caption2}
\usepackage{array}
\usepackage{fancyhdr}
\usepackage{cite,graphicx,amsmath,amssymb}
\usepackage{psfrag}
\usepackage{multirow}
\usepackage{longtable}

    \def\Complex{{\rm\rule[.23ex]{.03em}{1.1ex}\kern-.3em{C}}}

    \newcommand{\be}{\begin{equation}} \newcommand{\ee}{\end{equation}}
    \newcommand{\bea}{\begin{eqnarray}} \newcommand{\eea}{\end{eqnarray}}
    \newcommand{\benum}{\begin{enumerate}} \newcommand{\eenum}{\end{enumerate}}

\usepackage{color}

\ifCLASSINFOpdf

\else

\fi

\begin{document}
\title{A Survey of Physical Layer Security Techniques for 5G Wireless Networks and Challenges Ahead}

\author{Yongpeng Wu,~\IEEEmembership{Senior Member,~IEEE,}
Ashish Khisti,~\IEEEmembership{Senior Member,~IEEE,} \\Chengshan Xiao,~\IEEEmembership{Fellow,~IEEE,}
Giuseppe Caire,~\IEEEmembership{Fellow,~IEEE,} \\
Kai-Kit Wong,~\IEEEmembership{Fellow,~IEEE}, and Xiqi Gao ~\IEEEmembership{Fellow,~IEEE}

\thanks{Y. Wu is with Department of Electrical Engineering,  Shanghai Jiao Tong University,
Minhang, Shanghai, China 200240(Email:yongpeng.wu2016@gmail.com; yongpeng.wu@sjtu.edu.cn).}

\thanks{A. Khisti is with Signal Multimedia and Security Lab, University of Toronto, Bahen Building,
Toronto, Canada (Email: akhisti@ece.utoronto.ca).}

\thanks{C. Xiao is with Department of Electrical and Computer Engineering, Lehigh University, Bethlehem, PA 18015, USA (Email: xiaoc@lehigh.edu). }

\thanks{G. Caire is with Institute for Telecommunication Systems, Technical University Berlin, Einsteinufer 25,
10587 Berlin, Germany (Email: caire@tu-berlin.de). }

\thanks{K.-K. Wong is with Department of Electronic and Electrical Engineering, University College London,  London
WC1E 6BT, UK (Email:kai-kit.wong@ucl.ac.uk). }

\thanks{X. Q. Gao is with the National Mobile Communications Research Laboratory,
Southeast University, Nanjing, 210096, P. R. China (Email: xqgao@seu.edu.cn). }

}

\maketitle

\begin{abstract}
Physical layer security which safeguards
data confidentiality based on the information-theoretic approaches has received significant
research interest recently. The key idea behind physical layer security is to
utilize the intrinsic randomness of the transmission
channel to guarantee the security in physical layer.
The evolution towards 5G wireless communications poses new
challenges for physical layer security research. This paper provides
a latest survey of the physical layer security research on various
promising 5G technologies, including physical layer security coding,
massive multiple-input multiple-output, millimeter wave communications, heterogeneous networks, non-orthogonal multiple access,
full duplex technology, etc. Technical challenges which remain unresolved at the time of writing are
summarized and the future trends of physical layer security in 5G and
beyond are discussed.

\end{abstract}

\newpage

\section{Introduction}
Nowadays, wireless networks have been widely used in civilian and military applications
and become an indispensable part of our daily life. People rely heavily on wireless networks
for transmission of important/private information, such as credit card information,
energy pricing, e-health data, command, and control messages. Therefore, security
is a critical issue for future 5G wireless networks \cite{5G-PPP}.  Basically, the security today relies on bit-level cryptographic techniques
and associated protocols at various levels of the data processing stack. These solutions have drawbacks:
standardized protections within public wireless
networks are not secure enough, and many of their weaknesses are well known; even if enhanced ciphering and authentication protocols exist,
they occur strong constraints and high additional costs for the users of public networks, etc.
Therefore, new security approaches are issued from information theory fundamentals and focus on the secrecy capacity of the propagation channel,
which is referred as physical layer security \cite{Wyner1975BSTJ,Tekin2008TIT,Khisti2010TIT,Khisti2010TIT_2,Oggier2011TIT,Dong2011TSP,Fakoorian2011TSP,Zheng2011TSP,He2014TIT}.

The advantages of employing physical layer security techniques for 5G networks compared to
that of cryptography techniques are on two folds. First, physical layer security techniques do not rely on computational complexity.
As a result, even if the eavesdroppers (unauthorized smart devices) in the 5G networks are equipped with power computational devices,
the secure and reliable communications can still be achieved. In contrast, the security of computation-based
cryptography techniques will be compromised if the eavesdroppers' devices have sufficient computational
capacities for hard mathematical problem.
The 5G networks must simultaneously meet various different service requirements
with hierarchical architectures \cite{5G-PPP}, which implies that devices are always
connected to the nodes with different powers and different computational capacity levels.
Second, the structures of 5G networks are usually decentralized, which implies devices
may randomly connect in or leave the network at any time instants. For this case,
cryptographic key distribution and management become very challenging.
As a result, physical layer security techniques can be used
to either perform secure data transmission directly or generate
the distribution of cryptography keys in the 5G networks.
With careful management and implementation,
physical layer security can be used as an additional level
of protection  on top of the existing security schemes.
As such, they will formulate a well-integrated security solution together
that efficiently safeguards the confidential and privacy
communication data in 5G wireless networks.

Considering the potential of physical layer security for 5G wireless communications, the opportunities and challenges
on how the innovative 5G technologies achieve a high security level at the physical
layer deserves to receive more attention from the research community.
The purpose of this paper is to provide a comprehensive summarization of the latest physical layer security research results
on the key technologies of 5G wireless networks. In particular, we focus on the following typical technologies:

\begin{enumerate}

\item Physical layer security coding: Although the first
physical layer security code appears around 1970s,
to design explicit security codes which can be used in practical
communication systems is still challenging.
We review the state of art of three important physical layer security codes,
including low-density parity-check (LDPC) codes, polar codes, and lattice codes.

\item Massive MIMO: Deploying large antenna arrays significantly
increases the spatial dimension of wireless channels.  We discuss how to exploit
the extra spatial resources to effectively combat the eavesdropper
and guarantee the secure communication at physical layer. Both the
passive and the active eavesdropper scenarios are described.

\item Millimeter wave (mmWave) communications: Abundant
spectra within the high frequency band may
result in significant different propagation
environments for physical layer secure communication.
To understand mmWave secure transmission more clearly, research works
for both point-to-point and network mmWave communication systems are introduced.

\item Heterogeneous networks: In general, a heterogeneous network
is consisted of various tiers of networks which operates in
the same system bandwidth. We describe in details on
how to design transmission schemes
to secure multi-tier communications simultaneously.

\item Non-orthogonal multiple access (NOMA): As a multiple access technology, the security of NOMA communications
is an important concern which should be paid more attention. The physical layer security technology can
be combined with NOMA to tackle this issue.  The physical layer security of NOMA is a new and promising
research frontier, where a few relevant research results so far will be summarized.

\item Full duplex technology: The full duplex technology brings both the opportunity and the challenge for
the physical layer security communication. On one hand, the full duplex technology enables the receiver
to generate additional AN to interferer the eavesdropper. On the other hand,  the eavesdropper with
full duplex technology can actively attack the communication process while eavesdropping. In general, we
discuss four categorizations of full-duplex physical layer security communications,
including the full duplex receiver, the full duplex transmitter and receiver,
the full duplex base station, and the full duplex eavesdropper.

\end{enumerate}

Moreover, we also introduce other important results for physical layer security for future wireless networks,
such as the joint physical-application layer secure transmission design,
the practical test bed design for physical layer security, etc.
The future research challenges of physical layer security in 5G and
beyond are also discussed.

It should be noted that there are already many survey and tutorial papers
for physical layer security research \cite{Mukherjee2014CST,Yang2015CM,Zou2016PIEEE,Chen2017CST}. However, a comprehensive study
of physical layer security techniques for 5G wireless networks is still missing,
which is the main contribution of this paper. The most relevant work is \cite{Yang2015CM}, where
physical layer security for only three 5G techniques are briefly discussed in a big picture without
introducing the research results in details. In contrast, our paper provides a comprehensive detailed summarization
of latest research results on physical layer security for 5G wireless networks. Moreover,
the corresponding research challenges  at current stage are also discussed.

\section{Physical Layer Security Coding}
Most works on physical layer security are based on
non-constructive random-coding arguments to establish the theoretic results.
Such results demonstrate the existence of codes that achieve the secrecy
capacity, but are of little practical usefulness. The construction
of practical codes for physical layer security has received more attentions
recently. In this section, we review recent works on the construction of three practical codes for
physical layer security, which might be used in 5G communications.

\subsection{LDPC Codes}
In \cite{Thangaraj2007TIT}, A. Thangaraj \textit{et al.} establish
a connection between  capacity-achieving codes and secrecy based on the
metric of weak secrecy. It is proved in \cite{Thangaraj2007TIT}
that for an arbitrary wiretap channel, the perfect secrecy can
be achieved by using codes that achieve
the capacity of the eavesdropper's channel.
This conclusion provides an conceptual construction for designing
the secrecy transmission coding schemes over the general wiretap channel.
Moreover, the authors in \cite{Thangaraj2007TIT} use this idea to design
LDPC codes based on nested sparse graph codes and a coset coding scheme over a wiretap channel for a noiseless channel of the desire user
and a binary erasure channel (BEC) of the eavesdropper.
The constructed codes are the first secrecy-capacity-achieving LDPC codes in terms of
weak secrecy. Later, V. Rathi \textit{et al.} generalizes this coding scheme to BEC of both the desire user
and the eavesdropper by designing the two-edge type LDPC codes \cite{Rathi2013TIT}.
However, the proposed construction results in some degree
one variable nodes in the ensemble for the desire user's channel.
To circumvent this problem,  numerical  methods are used to optimize
the degree distribution of the two edge-type LDPC ensembles.
Some relatively simple ensembles are found, which achieve
good secrecy performance and are close to the secrecy capacity.
A. Subbramanian \textit{et al.} construct LDPC codes with large
girth block length based on Ramauja graph for a noiseless channel of the desire user
and BEC of the eavesdropper \cite{Subramanian2011TFS}, which achieve strong secrecy with lower rates.

LDPC codes have been designed for the Gaussian wiretap channel. The physical-layer security
communication is realized in \cite{Klinc2011TFS} by punctured LDPC codes under
the criterion of bit-error rate (BER), where the secrecy
information bits are hidden in the punctured bits. Therefore, these
information bits are not transmitted through the channel but can be decoded
at the receiver side based on the non-punctured part of the codeword.
This coding scheme can yield a BER close to 0.5 at the eavesdropper's side
while significantly reduces the security gap defined in \cite{Klinc2011TFS} comparing to the non-punctured LDPC codes.
However, the punctured LDPC codes result in higher power transmission comparing to the
non-punctured LDPC codes.
To solve this problem, M. Baldi \textit{et al.} propose a
nonsystematic coded transmission design
by scrambling the information bits \cite{Baldi2012TFS}. It is shown
in \cite{Baldi2012TFS} that this scrambling technique achieves
security gap comparable to that design based on puncturing but without
increasing the transmit power. This scrambling design has been extended
to parallel Rayleigh distributed channels \cite{Baldi2014TFS}.
By exploiting the equivocation rate of eavesdropper's channel as an optimization criterion,
M. Baldi \textit{et al.} propose a code design algorithm in the finite codeword length
regime \cite{Baldi2015}.  Based on this algorithm, irregular LDPC codes which approach the
ultimate performance limits with small codeword lengths are constructed.
A brief summary of above work is given in Table \ref{table:LPDC}.

\begin{table}
\centering
  \renewcommand{\multirowsetup}{\centering}
 \captionstyle{center}
  {
\caption{LDPC Codes for Physical Layer Security}
\label{table:LPDC}
\newsavebox{\tablebox}
\begin{lrbox}{\tablebox}
\begin{tabular}{|c|c|c|c|c|}
\hline
  Paper   & Main Channel & Eve Channel &  Criterion & Constituent Codes \\ \hline
A. Thangaraj \textit{et al.} \cite{Thangaraj2007TIT}   &  Noiseless  &  BEC & Weak secrecy & Duals of LDPC \\ \hline
V. Rathi \textit{et al.} \cite{Rathi2013TIT}   &  BEC  &  BEC & Weak secrecy & Two-edge LDPC \\ \hline
A. Subramanian \textit{et al.} \cite{Subramanian2011TFS}   &  Noiseless  &  BEC & Strong secrecy &  Duals of LDPC \\ \hline
D. Klinc  \textit{et al.} \cite{Klinc2011TFS}   &  Gaussian  &  Gaussian & BER & Punctured LDPC  \\ \hline
M. Baldi \textit{et al.} \cite{Baldi2012TFS}   &  Gaussian  &  Gaussian & BER & Non-punctured LDPC  \\ \hline
M. Baldi \textit{et al.} \cite{Baldi2014TFS}   &  Parallel Rayleigh  &   Parallel Rayleigh & BER & Non-punctured LDPC  \\ \hline
M. Baldi \textit{et al.} \cite{Baldi2015}   &  Gaussian  &   Gaussian & Equivocation rate of Eve &  Irregular LDPC  \\ \hline
\end{tabular}
\end{lrbox}
\scalebox{0.9}{\usebox{\tablebox}}
}
\end{table}

\subsection{Polar Codes}
For the  weak secrecy criterion, H. Mahdavifar \textit{et al.} construct a polar coding scheme
to achieve the secrecy capacity for the symmetric
binary-input memoryless wiretap channel under the condition that
the main channel of the eavesdropper is degraded
to the main channel of the desired user \cite{Mahdavifar2011TIT}.
The main idea in \cite{Mahdavifar2011TIT} is to select only those
bit channels which are good for both the desired user
and the eavesdropper to transmit random bits.
Moreover, those bit channels which are good for
desired user but bad for the eavesdropper are used
to transmit information bits. It is proved in \cite{Mahdavifar2011TIT} that
this coding scheme can achieve the secrecy capacity.
Furthermore, E. Hof \textit{et al.} and M. Andersson \textit{et al.}
 independently prove that this coding scheme achieves
the entire \textit{rate-equivocation region} (Defined in \cite{Liang2010FT})
in \cite{Hof2010} and \cite{Andersson2010CL}, respectively.
O. O. Koyluoglu \textit{et al.} apply this coding scheme
into a key agreement problem over the block fading wiretap channel \cite{Koyluoglu2012TFS}.
The secure polar code is used for each fading block, from which
the secrecy keys are generated based on standard privacy amplification techniques.
Y.-P. Wei \textit{et al.} develop polar codes for the general wiretap
channel by relaxing the degraded and the symmetric
constraints \cite{Wei2016JSAC}. In addition, this coding scheme is extended
to the multiple access wiretap channel (MA-WC), the broadcast channel with confidential
message (BC-CM), and the interference channel with confidential message (IC-CM).
On the other hand, S. A. A. Fakoorian \textit{et al.} design polar codes to achieve the secrecy capacity  for the arbitrary deterministic wiretap channel \cite{Fakoorian2013}. A polar coding scheme for bidirectional relay networks with common and confidential
messages and the decode-and-forward protocol is proposed in \cite{Andersson2013JSAC}.

For the strong security criterion,  E. Saso\v{g}lu \textit{et al.} design a multi-block polar coding
scheme \cite{Sasoglu2013} which achieves both security and reliability for the same wiretap
channel model as in \cite{Mahdavifar2011TIT}. T. C. Gulcu \textit{et al.}  provide a simple coding scheme
based on polar codes to achieve the secrecy capacity of the general wiretap channel
(not necessarily degraded or symmetric) \cite{Gulcu2015}. This coding scheme is also extended  to
achieve the capacity region of discrete memoryless BC-CM.
Independently, R. A. Chou \textit{et al.} design a more general (holds for more general conditions as given in \cite[Fig. 1]{Chou2016TIT})
polar coding scheme for discrete memoryless BC-CM \cite{Chou2016TIT}.  A brief summary of above work is given in Table \ref{table:Polar}.

Other polar coding schemes for wiretap channels include the concatenation of two polar codes for the general
wiretap channel \cite{Renes2013book}, the concatenation of polar and LDPC codes to minimize the security gap \cite{Zhang2014CL},
etc.

\begin{table}
\centering
  \renewcommand{\multirowsetup}{\centering}
 \captionstyle{center}
  {
\caption{Polar Codes for Physical Layer Security}
\label{table:Polar}
\begin{lrbox}{\tablebox}
\begin{tabular}{|c|c|c|c|}
\hline
  Paper   &  Channel &  Criterion  &    Main contribution \\ \hline
  H. Mahdavifar \textit{et al.} \cite{Mahdavifar2011TIT} & Symmetric
binary-input memoryless degraded wiretap channel    &  Weak secrecy & Achieve secrecy capacity    \\ \hline
 \cite{Hof2010,Andersson2010CL}  & Symmetric
binary-input memoryless degraded wiretap channel    &  Weak secrecy &  Achieve rate-equivocation region  \\ \hline
O. O. Koyluoglu \textit{et al.} \cite{Koyluoglu2012TFS}  & Symmetric
binary-input memoryless degraded wiretap channel    &  Weak secrecy &  Generate a key agreement   \\ \hline
\multirow{2}{*} {Y.-P. Wei \textit{et al.} \cite{Wei2016JSAC}}  & General wiretap channel   &  \multirow{2}{*} {Weak secrecy} &   Achieve secrecy capacity   \\
  &  MA-WC, BC-CM, IC-CM  &  &  Achieve secrecy rate regions  \\  \hline
S. A. A. Fakoorian \textit{et al.} \cite{Fakoorian2013}  & Deterministic wiretap channel   &  Weak secrecy &   Achieve secrecy capacity   \\ \hline
M. Andersson  \textit{et al.} \cite{Andersson2013JSAC}  & Bidirectional relay networks with confidential messages  &  Weak secrecy &   Achieve  capacity-equivocation region   \\ \hline
E. Saso\v{g}lu \textit{et al.} \cite{Sasoglu2013}  & Symmetric
binary-input memoryless degraded wiretap channel  &  Strong secrecy &   Achieves both security and reliability   \\ \hline
T. C. Gulcu \textit{et al.}  \cite{Gulcu2015}  & General wiretap channel   &  Strong secrecy &   Achieve secrecy capacity  \\ \hline
R. A. Chou \textit{et al.} \cite{Chou2016TIT}  & Discrete memoryless BC-CM  &  Strong secrecy &   Achieve  secrecy capacity  \\ \hline
\end{tabular}
\end{lrbox}
\scalebox{0.85}{\usebox{\tablebox}}
}
\end{table}

\subsection{Lattice Codes}
For wiretap lattice codes, J.-C. Belfiore \textit{et al.} define a notation of secrecy gain,
which reflects the eavesdropper's correct decoding probability \cite{Belfiore2010,Oggier2016TIT}.
Asymptotic analysis of the secrecy gain shows that it scales exponentially with the
dimension of the lattice. Also, examples of wiretap
lattice codes designed based on this secrecy gain criterion for the Gaussian wiretap channel are given in \cite{Oggier2016TIT}.
A.-M. E.-Hyt\"{o}nen proves that the symmetry points in the secrecy function
of the even extremal unimodular lattices achieve the secrecy gains \cite{Hytonen2012TIT}. In addition,
a method to examine the secrecy gains for arbitrary unimodular lattices is proposed
in \cite{Hytonen2012TIT}. F. Lin \textit{et al.} calculate
the symmetry points of four extremal odd unimodular lattices and 111 nonextremal
unimodular lattices for dimensions $8 < n \leq 23$. It is validated that these symmetry points are actually secrecy gains via
the method in \cite{Lin2013TIT}. Based on these secrecy gains, the best wiretap lattice codes are determined.
The lattice codes which are optimal based on the secrecy gain criterion for the Rayleigh fading wiretap channels
are designed in \cite{Belfiore2011}.

From information theory point of view, L.-C. Choo \textit{et al.} construct a nested lattice
code for the Gaussian wiretap channel based on the equivocation rate, which can meet both the reliability and the weak secrecy
criterions \cite{Choo2011}. C. Ling \textit{et al.} further design wiretap lattice codes
achieving the strong secrecy for the Gaussian wiretap channel \cite{Ling2014TIT}.  Moreover, L.-C. Choo \textit{et al.} propose a
superposition lattice code for the Gaussian BC with confidential message with the strong secrecy \cite{Choo2014}.
 A brief summary of above work is given in Table \ref{table:Lattice}.

\begin{table}
\centering
  \renewcommand{\multirowsetup}{\centering}
 \captionstyle{center}
  {
\caption{Lattice Codes for Physical Layer Security}
\label{table:Lattice}
\begin{lrbox}{\tablebox}
\begin{tabular}{|c|c|c|c|}
\hline
  Paper   &  Channel &  Criterion  &    Main contribution \\ \hline
 J.-C. Belfiore \textit{et al.} \cite{Belfiore2010,Oggier2016TIT} & Gaussian wiretap channel & Secrecy gain & Define secrecy gain  \\ \hline
A.-M. E.-Hyt\"{o}nen \textit{et al.} \cite{Hytonen2012TIT} & Gaussian wiretap channel & Secrecy gain & Propose a method to examine the secrecy gain  \\ \hline
F. Lin \textit{et al.} \cite{Lin2013TIT} & Gaussian wiretap channel & Secrecy gain & Construct best lattice codes for dimensions $8 < n \leq 23$  \\ \hline
J.-C. Belfiore \textit{et al.} \cite{Belfiore2011}  & Rayleigh wiretap channel & Secrecy gain & Construct a wiretap lattice code  \\ \hline
L.-C. Choo \textit{et al.} \cite{Choo2011}  & Gaussian wiretap channel & Weak secrecy  & Construct a nested  lattice code \\ \hline
C. Ling \textit{et al.} \cite{Ling2014TIT}  & Gaussian wiretap channel & Strong secrecy  & Design wiretap lattice codes \\ \hline
L.-C. Choo \textit{et al.} \cite{Choo2014}  & Gaussian BC  with confidential message & Strong secrecy  &  Propose a
superposition lattice code\\ \hline
\end{tabular}
\end{lrbox}
\scalebox{0.85}{\usebox{\tablebox}}
}
\end{table}

Other lattice code designs for the wiretap channel includes: nested lattices code designs for
cooperative jamming, interference channels, and the relay networks \cite{He2011TIT,He2013TIT,He2014TIT},
the security of the continuous mod-lattice channel with feedback \cite{Lai2008TIT}, etc.

\section{Physical Layer Security in Massive MIMO Systems}
Massive MIMO is a promising approach for efficient transmission of massive information
and is regarded as one of ``big three" 5G technologies \cite{Andrews2014JSAC}.
In this section, we review the current
security threats and countermeasures of massive MIMO technology
based on passive and active eavesdropper scenarios, respectively.

\subsection{Passive Eavesdropper Scenarios}
Physical layer security for massive MIMO systems with
passive eavesdroppers has been recently studied.
J. Zhu \textit{et al.} study secure massive MIMO transmissions for multi-cell multi-user systems over
i.i.d. Rayleigh fading channel \cite{Zhu2014TWC}, where a passive eavesdropper attempts to decode the information sent to one of the users.
The impact of multi-cell interference and pilot contamination on the achievable erogdic secrecy rate
are analyzed and several matched filtering precoding and artificial noise (AN) generation designs are proposed
to degrade the eavesdropper's channel and protect the desired user's channel.
For the same system model, regularized channel inversion and AN transmission schemes
are designed in \cite{Zhu2016TWC} to further improve the secrecy rate performance.
J. Wang \textit{et al.} investigate AN-aided secure massive MIMO transmission
over i.i.d. Rician fading channel \cite{Wang2015TCom}.
For single-cell multiuser massive MIMO systems with
distributed antennas, K. Guo \textit{et al.} design
three secure-constrained power allocation schemes \cite{Guo2016TWC} by maximizing the minimum user's signal-to-interference-noise ratio (SINR)
subject to the eavesdropper's SINR and the sum power constraint and minimizing the sum transmit power
subject to SINR constraints of users and the eavesdropper, respectively.
Y. Wu \textit{et al.} investigate secure transmission designs for large-scale MIMO systems
with finite alphabet inputs \cite{Wu2017}. Power allocation schemes for relay-aided large-scale
MIMO systems are proposed in \cite{Chen2015TWC,Chen2016TFS}.
A brief summary of above work is given in Table \ref{table:massive_passive}.

\begin{table}
\centering
  \renewcommand{\multirowsetup}{\centering}
 \captionstyle{center}
  {
\caption{Secure Massive MIMO with Passive Eavesdropper}
\label{table:massive_passive}
\begin{lrbox}{\tablebox}
\begin{tabular}{|c|c|c|}
\hline
  Paper   & System Model &  Main Contribution  \\ \hline
 J. Zhu \textit{et al.} \cite{Zhu2014TWC}   &  Multi-cell multi-user, one desired user, one eavesdropper, i.i.d. Rayleigh &  Matched filtering precoding and AN generation designs  \\ \hline
 J. Zhu \textit{et al.} \cite{Zhu2016TWC}   &  Multi-cell multi-user, one desired user, one eavesdropper, i.i.d. Rayleigh &  Regularized channel inversion and AN generation designs  \\ \hline
 J. Wang \textit{et al.}  \cite{Wang2015TCom}   & One desired user, multiple eavesdroppers, i.i.d. Rician & AN-aided secure transmission designs   \\ \hline
  K. Guo  \textit{et al.}  \cite{Guo2016TWC}   & Single-cell, multiple desired users, one eavesdropper, correlated Rayleigh & Distributed power allocation under security-constraints \\ \hline
    Y. Wu  \textit{et al.}  \cite{Wu2017}   & One desired user, one eavesdropper, perfect CSI & Secure transmission with finite alphabet inputs \\ \hline
X. Chen \textit{et al.} \cite{Chen2015TWC,Chen2016TFS} &  Relay-aided, one desired user, one eavesdropper, i.i.d. Rayleigh &  Secrecy performance analysis and power allocation designs\\ \hline
\end{tabular}
\end{lrbox}
\scalebox{0.85}{\usebox{\tablebox}}
}
\end{table}

Other secure massive MIMO work with passive eavesdroppers include:
secure transmission for massive MIMO systems with limited radio frequency and hardware impairments \cite{Zhu2017,ZhuJ2017TWC},
secure strategies in presence of a massive MIMO eavesdropper \cite{Chen2016IA,Chen2016IA_2},
secrecy outage probability analysis for massive MIMO systems \cite{Wei2015}, etc.

\subsection{Active Eavesdropper Scenarios}
Most physical layer security research work assume that perfect channel knowledge
of the legitimate user is available at the transmitter and do not consider the procedure required
to obtain this channel.
In time duplex division (TDD) communication systems,
the users in an uplink training phase will send
pilot signals to the base station (BS)
to estimate the channel for
the subsequent downlink transmission.
From the eavesdropper's point of view, it can
 actively send the same pilot signals as the users
to attack this uplink channel training phase and hence significantly increase its eavesdropping capability \cite{Zhou2012TWC}.

\begin{figure*}[!ht]
\centering
\includegraphics[width=0.7\textwidth]{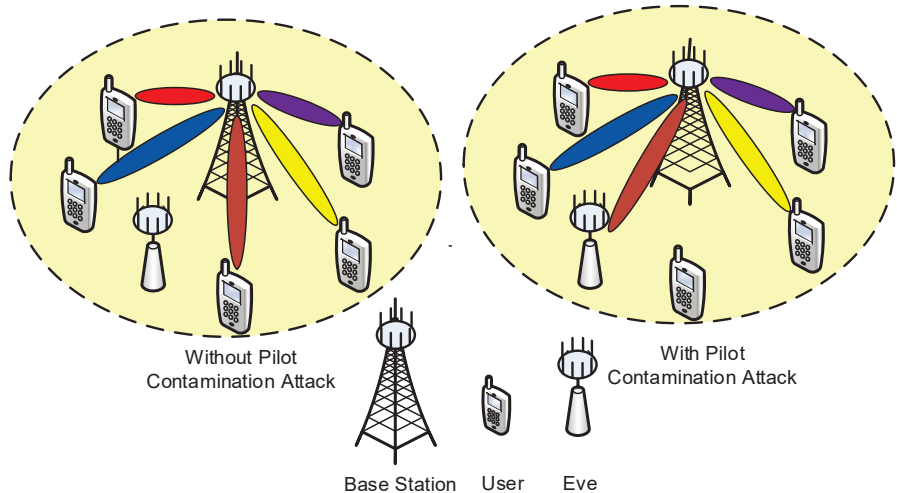}
\caption {\space\space Pilot contamination attack on massive MIMO systems}
\label{Pilot_Contamination}
\end{figure*}

\begin{figure*}[!ht]
\centering
\includegraphics[width=0.7\textwidth]{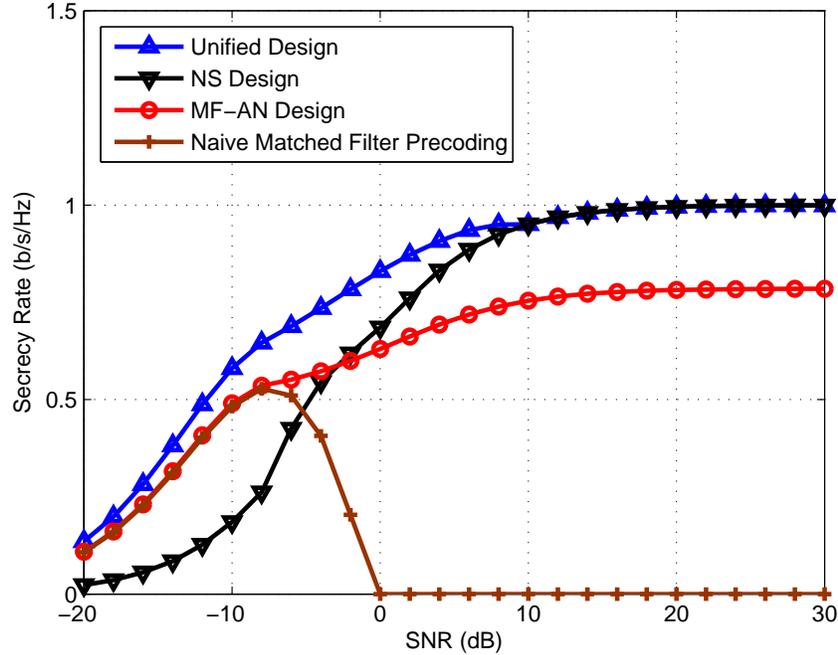}
\caption {\space\space Exact secrecy rate vs. the SNR for 128 antennas massive MIMO systems and different precoding designs.
Experimental results extracted from \cite{Wu2016TIT}.}
\label{Secure_Massive MIMO}
\end{figure*}

This pilot contamination attack causes a serious secrecy threat to TDD-based massive MIMO systems.
On one hand, large antenna arrays beamforming leads to
the hardening of the channel, which prevents
the exploitation of channel fluctuations caused by
fading  to improve the secrecy performance.
On the other hand, as illustrated in Figure \ref{Pilot_Contamination},
the pilot contamination attack enables  the transmitter to  beamform
towards the the eavesdropper instead of the desired user.
If the eavesdropper's pilot power is sufficiently large, a positive secrecy rate
may not be achievable.  This is significantly different from
the conventional idea that massive MIMO naturally facilitates secure communication
since the large antenna arrays can generate very narrow beams focusing on the desired users without
spilling over the signal power in other directions.
In the first time, Y. Wu \textit{et al.} systematically analyze the secrecy threat caused by
the pilot contamination attack for multi-cell multi-user massive MIMO systems over correlated fading channels \cite{Wu2016TIT}.
Then, a matched filter precoding and AN generation design
and a null space design are provided in \cite{Wu2016TIT} to combat the pilot contamination attack for weakly correlated channels
and highly correlated channels, respectively.
A unified design which combines the matched filter precoding and AN generation design
and the null space design is also proposed.
Simulations indicate that these designs can guarantee
reliable secure communication under the pilot contamination attack, as shown in Fig. \ref{Secure_Massive MIMO}.
At the same time, Y. O. Basciftci \textit{et al.} study
the pilot contamination attack problem for single-cell multi-user massive MIMO systems
over i.i.d. fading channels \cite{Basciftci2017}. It is proved in \cite{Basciftci2017} that if the pilot contamination attack
does not exist, the maximum secure degree of freedom (DoF) of massive MIMO systems
is the same as the maximum DoF of massive MIMO systems when the eavesdropper does not exist.
However, if the pilot contamination attack exists, the maximum secure DoF of massive MIMO systems
could be zero. To defense the pilot contamination attack,
Y. O. Basciftci \textit{et al.} expand cardinality of the pilot signal set and hide the
pilot signal within the enlarged set. On the other hand,
S. Im \textit{et al.} employ a secret key agreement
protocol for single-cell multi-user massive MIMO systems with
the pilot contamination attack \cite{Im2015TWC}.
An estimator is designed at the BS side to evaluate the information leakage.
Then, the BS and the desired user perform the
reliable secure communication by adjusting the lengths
of the secrecy key based on the estimated information leakage.
A brief summary of above work is given in Table \ref{table:massive_active}.

\begin{table}
\centering
  \renewcommand{\multirowsetup}{\centering}
 \captionstyle{center}
  {
\caption{Secure Massive MIMO with Active Eavesdropper}
\label{table:massive_active}
\begin{lrbox}{\tablebox}
\begin{tabular}{|c|c|c|}
\hline
  Paper   & System Model &  Main Contribution  \\ \hline
\multirow{2}{*} {Y. Wu \textit{et al.} \cite{Wu2016TIT}}    &  Multi-cell multi-user, one desired user & Systematically analyze the secrecy threat caused by
the pilot contamination attack   \\
& one eavesdropper, correlated Rayleigh   & Propose efficient schemes to combat the pilot contamination attack \\ \hline
\multirow{2}{*} {Y. O. Basciftci \textit{et al.} \cite{Basciftci2017}}    &  Single-cell multi-user, multiple desired users& Analyze maximum secure DoF with the pilot contamination attack  \\
& one eavesdropper, i.i.d. Rayleigh  & Propose a scheme to hide the pilot signal under the pilot contamination attack  \\ \hline
\multirow{2}{*} {S. Im \textit{et al.} \cite{Im2015TWC}}    &  Single-cell multi-user, one desired user & Employ
secret key agreement protocol with the pilot contamination attack  \\
& one eavesdropper, i.i.d. Rayleigh  & Adjust the lengths
of the secrecy key based on the estimated information leakage  \\ \hline
\end{tabular}
\end{lrbox}
\scalebox{0.85}{\usebox{\tablebox}}
}
\end{table}

Other secure massive transmission against active eavesdropper includes pilot retransmission strategies \cite{Do2017}
and the secure transmission design based on game theory \cite{Rawat2016Info}, etc.

\section{Physical Layer Security for mmWave Communications}
One of the most promising potential 5G technologies under consideration is the use
of high-frequency signals in the millimeter-wave frequency band that could allocate more bandwidth to deliver faster, higher-quality video and multimedia content \cite{Rappaport2013AC}.
Comparing to micro-Wave networks, the mmWave networks have various new characteristics such as the large number of antennas, short range
and highly directional transmissions, different propagation laws, and sensitive to blockage effects, etc. Therefore, the secure mmWave communications
will be different from the conventional secure micro-Wave communications.

L. Wang \textit{et al.} first show that the high secrecy throughout can
be achieved for a point-to-point mmWave communication system with multiple eavesdroppers \cite{Wang2014SPAWC}.
Assuming the uniform linear array (ULA) at the transmitter,
an analog beamforming with phase shift is employed based on the perfect CSI of the desired user.
The ergodic secrecy rate expressions are derived for both delay-tolerant and delay-limited transmission modes.
In particular, simulations show that with a large number of
transmit antennas, the delay-tolerant transmission mode
can achieve multi-gigabit per second secrecy rate at mmWave frequencies.
Motivated by this, Y. Ju \textit{et al.} further evaluate the secrecy performance
of the mmWave communication over the
multiple-input, single-output, single-antenna eavesdropper (MISOSE) wiretap
channel \cite{Ju2017TCOM}. Based on the perfect CSI of the desired user
and the statistical CSI of the eavesdropper, the secrecy
outage probability and secrecy throughout of the
matched filter precoding and AN generation design are analyzed.
The obtained results reveal that
the overlap between the desired user's and the eavesdropper's spatially resolvable paths has
an significantly important impact on the secrecy performance of the mmWave communication.
X. Tian \textit{et al.} investigate the hybrid precoder design for mmWave
multiple-input, multiple-output, multiple-antenna eavesdropper (MIMOME) wiretap channel \cite{Tian2017}.

\begin{figure*}[!ht]
\centering
\includegraphics[width=0.7\textwidth]{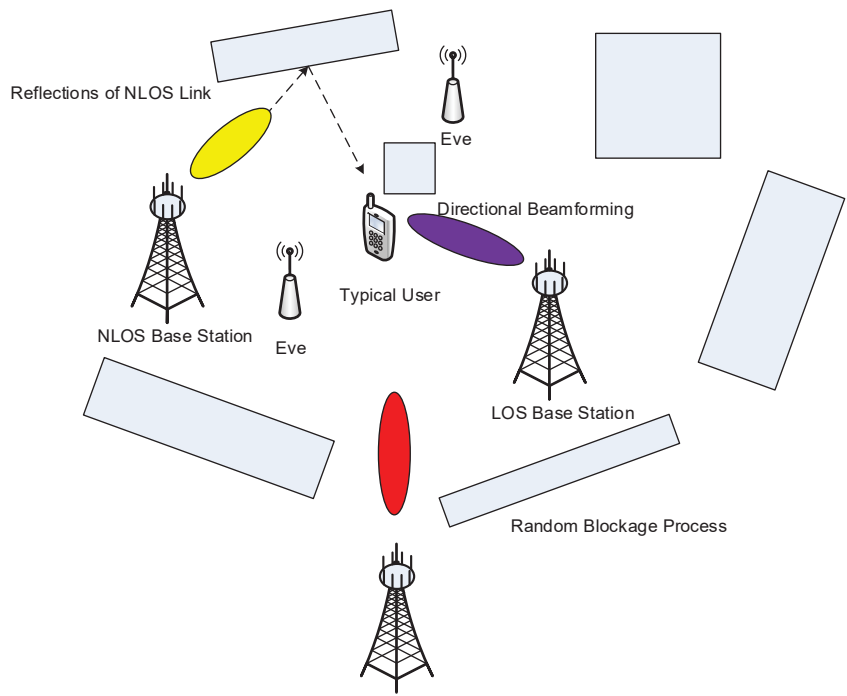}
\caption {\space\space Downlink secure communication for the mmWave cellular network}
\label{mmWave_systems}
\end{figure*}

\begin{figure*}[!ht]
\centering
\includegraphics[width=0.7\textwidth]{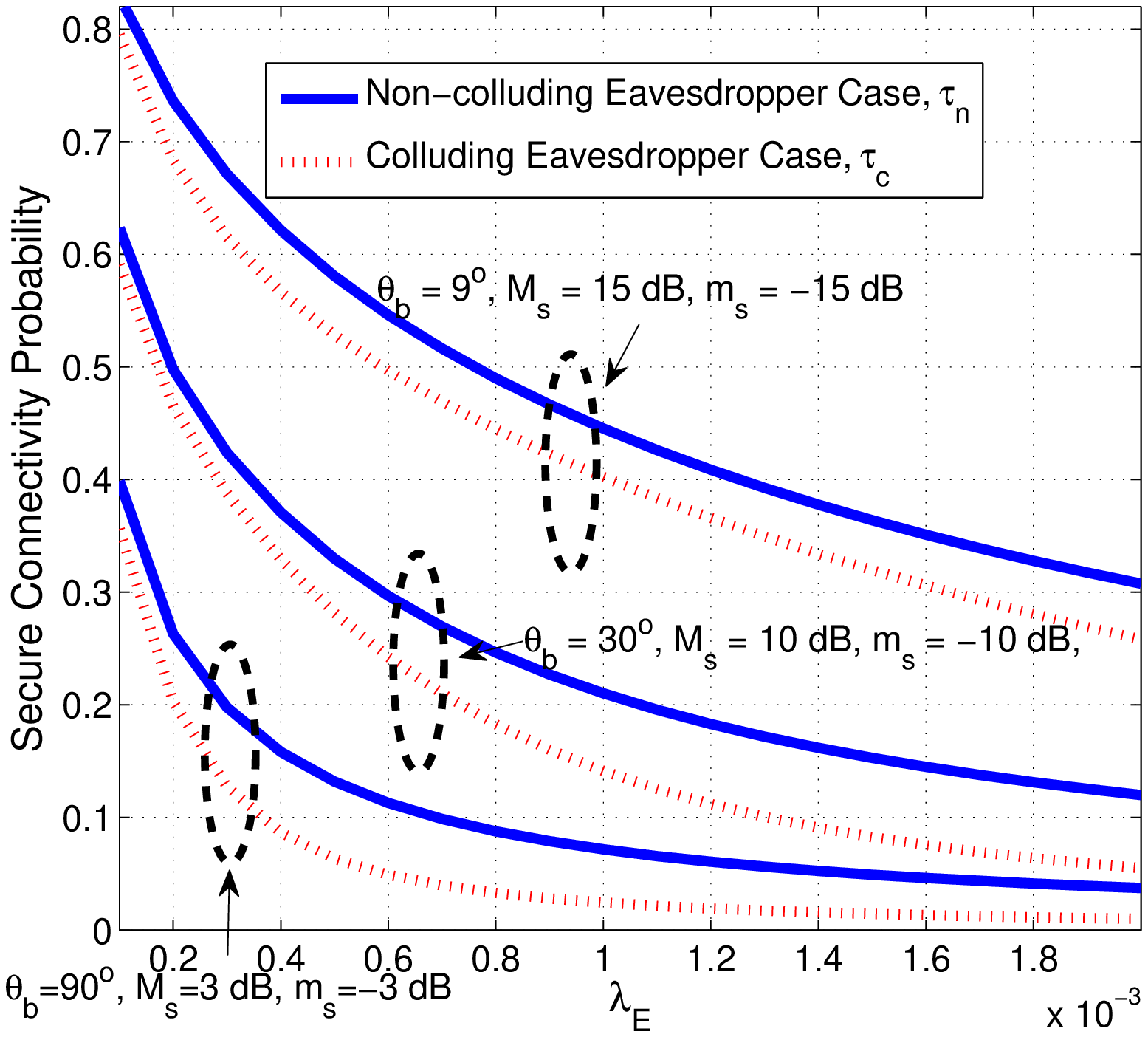}
\caption {\space\space Secrecy connectivity probability of mmWave cellular networks. Experimental results extracted
from \cite{Wang2016TWC}. $\theta_b$, $M_s$, $m_s$ denote the beam width of the main lobe, the array gain of the main lob, and the array gain
of the sidelobe of the directional beamforming, respectively. $\lambda_{E}$ denote the intensity of eavesdroppers.}
\label{SectorGain}
\end{figure*}

Based on the stochastic geometry framework,
C. Wang \textit{et al.} investigate
the downlink secure communication for mmWave cellular networks as shown in Fig. \ref{mmWave_systems} \cite{Wang2016TWC}.
The BSs perform the directional beamforming with the intended users' perfect CSI
and both the legitimate users and eavesdroppers in the networks
are equipped with a single omnidirectional antenna. As indicated in Fig. \ref{SectorGain},
it is shown that narrowing the beam width of directional beamforming antenna with more focused array gain
is beneficial for increasing the secrecy performance of mmWave networks.
In addition, the effects of antenna array pattern, base station intensity,
and AN generation on secrecy performance are investigated in \cite{Wang2016TWC}.
For the same system model, S. Vuppala \textit{et al.} further
analyze the secrecy performance for the mmWave and the micro-Wave hybrid
communication \cite{Vuppala2016TCOM}. It is revealed in \cite{Vuppala2016TCOM} that the blockages in mmWave
networks can decrease the secrecy outage probability.

Y. Zhu \textit{et al.} investigate the physical layer security for large-scale mmWave ad hoc networks,
which are modeled based on stochastic geometry \cite{Zhu2017TWC}. The directional beamforming is used between the transmitters
and the desired receivers and the corresponding average secrecy rate is derived. For the special case of
ULA, an explicit expression for the average secrecy rate is obtained, which reveals
that using more antennas at the transmitters is beneficial for suppressing the array gains
at the eavesdroppers' side. The proper power allocation between the transmit signal and
AN is also discussed. S. Gong \textit{et al.} investigate the secure precoding design
for mmWave two-way amplify-and-forward (AF) MIMO relaying networks by exploiting
the global perfect CSI \cite{Gong2017TVT}.
To reduce the hardware cost and power consumption for the mmWave communication,
an additional rank constraint is posed on the precoding matrix at the relay
to control the number of analog-to-digital converters in the system.

The mmWave communication system is usually equipped with a large number of
antennas at the transmitter with a limited number of radio frequency (RF) chains.
To take advantage of this point, N. Valliappan \textit{et al.}
consider another approach by using an antenna subset modulation (ASM) technique
to reach secure mmWave communication at physical layer \cite{Valliappan2013TCOM}.
The proposed approach utilizes a subset of antenna array to formulate a directional
modulation signal intended for the desired user.
By randomly choosing the antenna subset for each symbol, the received signal
for the undesired user becomes a randomized noise. Therefore, the
secure transmission is achieved. M. E. Eltayeb \textit{et al.} further
extended this ASM technique to mmWave vehicular communication systems \cite{Eltayeb2017}.
A brief summary of above work is given in Table \ref{table:mmWave}.

From the eavesdropper's point of view, the work in \cite{Steinmetzer2015} also
considers the potential attack in mmWave communication.
It is shown in \cite{Steinmetzer2015} that even by using highly directional mmWaves,
the reflection signals caused by small-scale physical objects will
be beneficial to the eavesdropper. In some cases, the achievable secrecy rate
can be decreased to zero by the attack.

\begin{table}
\centering
  \renewcommand{\multirowsetup}{\centering}
 \captionstyle{center}
  {
\caption{Secure mmWave Communications}
\vspace{0.2cm}
\label{table:mmWave}
\begin{lrbox}{\tablebox}
\begin{tabular}{|c|c|c|c|}
\hline
  Paper   & System Model & CSI  & Objective   \\ \hline
 \multirow{2}{*} {L. Wang \textit{et al.} \cite{Wang2014SPAWC}} & Point-to-point MISO & Perfect CSI of the desired user &  \multirow{2}{*} {Analyze ergodic secrecy rate} \\
 & Multiple single-antenna eavesdroppers   & Statistical CSI of eavesdroppers &  \\ \hline
 \multirow{2}{*} {Y. Ju \textit{et al.} \cite{Ju2017TCOM}}  &  \multirow{2}{*} {MISOSE wiretap channel} & Perfect CSI of desired user &  Analyze secrecy outage probability \\
 &    & Statistical CSI of the eavesdropper & Analyze secrecy throughout  \\ \hline
  \multirow{2}{*} {X. Tian \textit{et al.}  \cite{Tian2017}}  &  \multirow{2}{*} {MIMOME wiretap channel} & Perfect CSI of desired user &  \multirow{2}{*} {Hybrid precoders design}  \\
 &    & Perfect CSI/No CSI of the eavesdropper &  \\ \hline
   \multirow{2}{*} {C. Wang \textit{et al.}  \cite{Wang2016TWC}}  &     \multirow{2}{*} {Stochastic geometry cellular networks} & Perfect CSI of desired user &  Analyze secure connectivity probability  \\
 &    & Statistical CSI of the eavesdropper &  Evaluate the effect of AN   \\ \hline
    \multirow{2}{*} {S. Vuppala \textit{et al.} \cite{Vuppala2016TCOM}}  &  Stochastic geometry cellular networks & Perfect CSI of the desired user &
   \multirow{2}{*} {Analyze the secrecy outage probability} \\
 & MmWave/micro-Wave hybrid communication   & Statistical CSI of the eavesdropper &    \\ \hline
     \multirow{2}{*} {Y. Zhu \textit{et al.} \cite{Zhu2017TWC}}  &  \multirow{2}{*} {Stochastic geometry mmWave ad hoc networks} & Perfect CSI of desired user &
   \multirow{2}{*} {Analyze the average secrecy rate} \\
 &   & Statistical CSI of the eavesdropper &    \\ \hline
     \multirow{2}{*} {S. Gong \textit{et al.} \cite{Gong2017TVT}}  & AF MIMO relay networks & \multirow{2}{*} {Global perfect CSI} &
   \multirow{2}{*} {Secrecy precoders design } \\
 & A multiple-antenna eavesdropper  &  &    \\ \hline
 \multirow{2}{*} {N. Valliappan \textit{et al.}  \cite{Valliappan2013TCOM}}       & MISO, limited RF chains&  Perfect CSI of the desired user &
   \multirow{2}{*} {Formulate directional modulation} \\
 &  Single-antenna eavesdroppers  & No CSI of the eavesdropper &    \\ \hline
  \multirow{2}{*} {M. E. Eltayeb \textit{et al.} \cite{Eltayeb2017}}       & MISO, a single RF chain (vehicular systems)&  Perfect CSI of the desired user &
   \multirow{2}{*} {Formulate directional modulation} \\
 &  A single-antenna eavesdropper  & No CSI of the eavesdropper &    \\ \hline
\end{tabular}
\end{lrbox}
\scalebox{0.85}{\usebox{\tablebox}}
}
\end{table}

\section{Physical Layer Security in Heterogeneous Networks}
\begin{figure*}[!ht]
\centering
\includegraphics[width=0.7\textwidth]{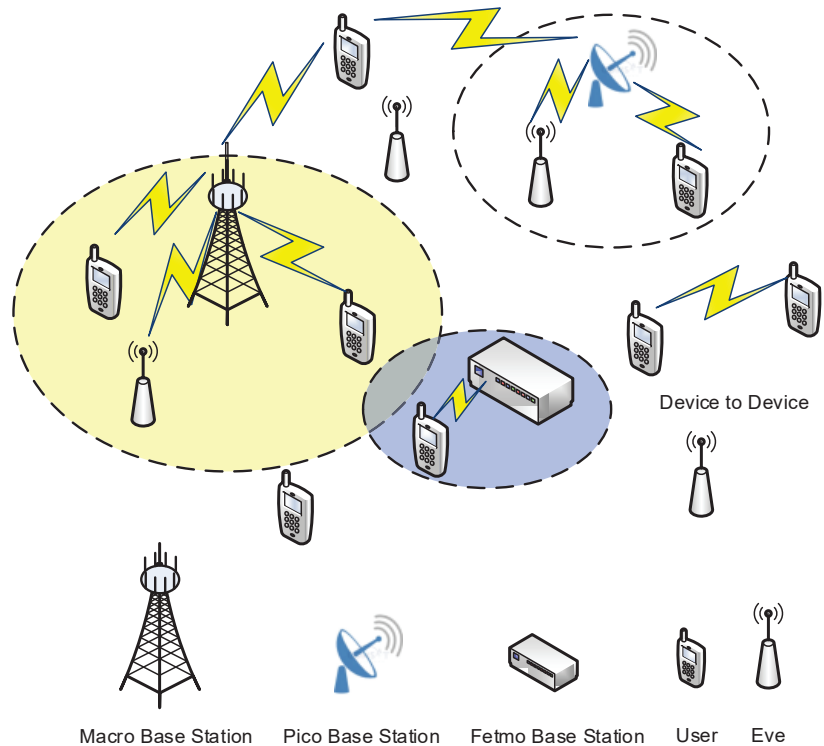}
\caption {\space\space A 4-tier macro/pico/femto/D2D heterogeneous network with users and eavesdroppers.}
\label{HCN}
\end{figure*}

The 5G heterogeneous networks should intelligently and seamlessly integrate
multiple nodes to form a multi-tier hierarchical architecture,
including the \emph{macro cell tiers} with high-power nodes for large radio coverage areas,
the \emph{small cell tiers} with low-power nodes for small radio coverage areas, the \textit{device tiers}
which support device to device communications, etc.
Fig. \ref{HCN} shows a typical 4-tier macro/pico/femto/D2D
heterogeneous network with users and eavesdroppers.
This multi-tier architecture brings new challenges to
the investigation of physical layer security compared to the conventional single-tier topology.
For example, the locations of the high/low power nodes will have a significant impact on
the physical layer security design, which need to be modeled and analyzed properly.
The optimal selection policy for each user among high/low power nodes under security constraints becomes difficult.
The protection of confidential and privacy data between connected devices against data leakage requires sophisticated
designs. Moreover, heterogeneous networks may introduce severe cross-tier interference. This
 should be taken into consideration when designing the reliable and secure data transmission schemes.
 In addition, users are accessible to an arbitrary tier, e.g., open access. Therefore, specific user association policies
 that coordinate both quality of service and secrecy are necessary.

T. Lv \textit{et al.} first study the physical layer security
in a downlink two-tier heterogeneous network with multiple single-antenna
users and a single-antenna eavesdropper in each cell \cite{Lv2015JSAC}.
Both the orthogonal spectrum allocation (OSA) scheme  and the secrecy-oriented non-orthogonal spectrum allocation (SONOSA)
scheme are considered. For OSA scheme, no interference
from other cells exists and a transmit strategy to maximize
the secrecy rate of one desired user under the quality of service (QoS)
constraints of other users is proposed. For SONOSA, some femtocell base stations
near the eavesdropper are allocated the same spectrum efficiency as the macrocell base station.
Then, these femtocell base stations cooperate to generate the maximal interference
at the eavesdropper side while guaranteeing the QoS requirements for the femtocell users.
H. Wu \textit{et al.} propose a user association policy by comparing
the average received signal at
the desired user with a given threshold \cite{Wu2016CLett}.
Then, assuming both the inter-and intra-cell interference do not exist,
closed-form expressions of
secrecy outage probability are obtained for a downlink K-tier heterogeneous network with single-antenna nodes by modeling
the locations of the nodes as independent Poisson Point Processes (PPPs).
Furthermore, Y. J. Tolossa \textit{et al.} derive
the average secrecy rate expression under the association policy that
any potential base station who provides $k$th largest path gain to the user
can be selected as the association base station \cite{Tolossa2017IOT}.
For a downlink K-tier heterogeneous network with only intra-cell interference,
M. Xu \textit{et al.} propose a dynamic coordinated multipoint transmission (CoMP)
scheme to increase the secure communication coverage \cite{Xu2016CLett}.

By considering both inter-and intra-cell interference, H.-M. Wang \textit{et al.} obtain
both secrecy and connection probabilities of a downlink K-tier heterogeneous network
based on a truncated average received signal power user association policy \cite{Wang2016TCOM}. A trade
off between secrecy and connection probabilities
in terms of association threshold, base station density, and power allocation
between the useful signal and AN is revealed. The network-wide secrecy throughout
and minimum secrecy throughout per user subject to both secrecy and connection probability constraints
are further analyzed. W. Wang \textit{et al.} derive closed expressions of
secrecy and connection probabilities for a downlink small cell network \cite{Wang2017TFS}.
The obtained results indicate that increasing the base station density is beneficial
for both secrecy and connection outage probability performance.
The performance of a uplink two-tier heterogeneous
network is investigated in \cite{Wu2016Globecom}. By using two-dimensional PPPs to approximate the summation of all interference
in the whole space, the authors derive the exact expressions for the successful connection and
secrecy outage probabilities. The secure transmission with wireless information and power transfer in a downlink two-tier
heterogeneous network is studied in \cite{Ren2017Access}. A brief summary of above work is given in Table \ref{table:HCN}.

\begin{table}
\centering
  \renewcommand{\multirowsetup}{\centering}
 \captionstyle{center}
  {
\caption{Secure transmission in heterogeneous networks}
\label{table:HCN}
\begin{lrbox}{\tablebox}
\begin{tabular}{|c|c|c|}
\hline
  Paper   & System Model &  Main Contribution  \\ \hline
T. Lv \textit{et al.} \cite{Lv2015JSAC}    & Downlink two-tier heterogeneous network & Propose secure transmission schemes  in the first time  \\
\hline
H. Wu \textit{et al.} \cite{Wu2016CLett}    & Downlink K-tier heterogeneous network, no interference& Analyze secrecy outage probability  \\
\hline
Y. J. Tolossa \textit{et al.} \cite{Tolossa2017IOT}    & Downlink K-tier heterogeneous network, no interference& Analyze average secrecy rate   \\
\hline
M. Xu \textit{et al.} \cite{Xu2016CLett}    & Downlink K-tier heterogeneous network, intra-cell interference & Incorporate a dynamic CoMP scheme  \\
\hline
H.-M. Wang \cite{Wang2016TCOM}    & Downlink K-tier heterogeneous network, inter-and intra-cell interference & Analyze secrecy and connection probabilities  \\
\hline
W. Wang \cite{Wang2017TFS}    & Downlink small network, inter-and intra-cell interference & Analyze secrecy and connection probabilities  \\
\hline
H. Wu \cite{Wu2016Globecom}    & Uplink two-tier heterogeneous network, inter-and intra-cell interference & Analyze secrecy and connection probabilities  \\
\hline
   \multirow{2}{*} {Y. Ren \cite{Ren2017Access}}    & Downlink two-tier heterogeneous network & \multirow{2}{*} {Secure transmission design}  \\
    & Wireless information and power transfer &    \\
\hline
\end{tabular}
\end{lrbox}
\scalebox{0.85}{\usebox{\tablebox}}
}
\end{table}

\section{Physical Layer Security of NOMA}
Given the scarce bandwidth resource, NOMA plays a crucial role for providing large system throughput,
high reliability, improved coverage, low latency, and massive connectivity in 5G wireless networks \cite{Huawei}.
As a result,  NOMA has been recognized as an important enabling technology in 5G wireless communication systems.
Because of the spectral efficiency benefit, NOMA has been recently included in 3GPP long term evolution advanced (LTE-A),
which further evidences the importance of NOMA in future wireless networks.
Thus, providing an unrivalled level of security for NOMA technology is one of the top priorities in the design
and implementation of the 5G wireless networks.  A significant effort is needed to efficiently combine physical layer
security with NOMA.  However, some challenges need to be resolved during the design process, such as the dissimilar
transmit powers and heterogeneous security requirements of users.  In addition, cooperation among users offers an interesting
option to enhance the secrecy performance.

\begin{figure*}[!ht]
\centering
\includegraphics[width=0.7\textwidth]{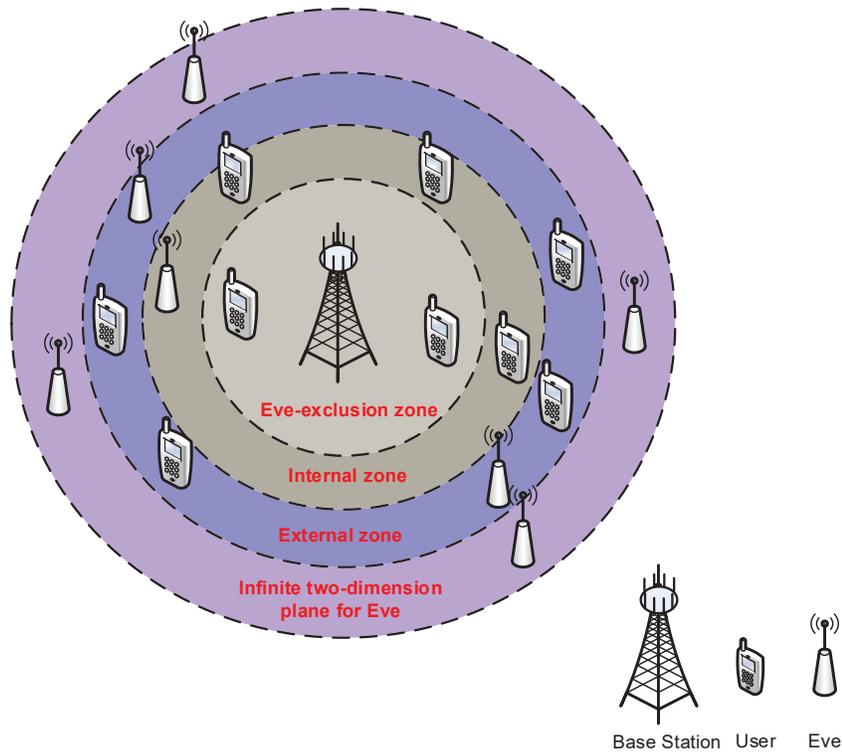}
\caption {\space\space Network model for the secure NOMA transmission.}
\label{NOMA}
\end{figure*}

Y. Zhang \textit{et al.} first study the secure NOMA transmission
for a single-antenna, one transmitter, multiple users, and one
eavesdropper system with perfect CSI of the users and no CSI of the eavesdropper \cite{Zhang2016CLetter}.
Perfect successive interference cancellation (SIC) is performed at each user.
The closed-formed expression of the optimal power allocation policy which maximizes the secrecy sum rate under each user's QoS constraint
is derived. For the same CSI assumption, Y. Liu \textit{et al.} further investigate
the secure transmission for NOMA networks \cite{Liu2017TWC}, as shown in Fig. \ref{NOMA}.
An eavesdropper-exclusion zone is established.
To reduce the SIC complexity at the receiver, a user paring scheme
is employed, where one user in the internal zone and one user in the external
zone are allocated the same resource slot. When the base station only has
single antenna, the secrecy outage probability is analyzed. Moreover,
the secrecy diversity order is obtained, which reveals that the user
with the weaker channel in the pair determines the secrecy diversity order.
When the base station is equipped with multiple antennas,
a matched filter precoding and AN generation design is employed to further increase
the secrecy performance. Based on this design, both the exact and asymptotic (in large system limit)
secrecy outage probability expressions are derived.
With perfect CSI and perfect SIC assumption, Z. Ding \textit{et al.} consider a NOMA network with both multicasting and
unicasting transmissions \cite{Ding2017TCom}. For the unicasting transmission, it it proved that the secrecy rate for NOMA
is no less than that for orthogonal multiple access in the high SNR regime. Also, the
secrecy outage probability is studied.  A brief summary of above work is given in Table \ref{table:NOMA}.

\begin{table}
\centering
  \renewcommand{\multirowsetup}{\centering}
 \captionstyle{center}
  {
\caption{Physical Layer Security of NOMA}
\vspace{0.2cm}
\label{table:NOMA}
\begin{lrbox}{\tablebox}
\begin{tabular}{|c|c|c|c|}
\hline
  Paper   & System Model & CSI  & Objective   \\ \hline
 \multirow{2}{*} {Y. Zhang \textit{et al.} \cite{Zhang2016CLetter}} & Single-antenna nodes & Perfect CSI of users &  \multirow{2}{*} {Maximize secrecy sum-rate} \\
 & A transmitter, multiple users, and a
eavesdropper & No CSI of the eavesdropper &  \\ \hline
 \multirow{2}{*} {Y. Liu \textit{et al.} \cite{Liu2017TWC}} & Downlink single-antenna networks & Perfect CSI of users &  \multirow{2}{*} {Analyze secrecy outage probability} \\
 & Downlink multiple transmit antenna networks & No CSI of the eavesdropper &  \\ \hline
  \multirow{2}{*} {Z. Ding \textit{et al.}\cite{Ding2017TCom}} & Downlink MISO networks with  & \multirow{2}{*} {Global perfect CSI}  &  Analyze secrecy outage probability  \\
 & multicasting and unicasting transmissions &  &  for the unicasting transmission   \\ \hline
\end{tabular}
\end{lrbox}
\scalebox{0.9}{\usebox{\tablebox}}
}
\end{table}

\section{Physical Layer Security for Full Duplex Technology}
Transmitting information by full duplex technology
consists in transmitting and receiving simultaneously on the same frequency band.
Theoretically, full duplex communications can double the spectral efficiency compared to
the conventional half duplex communications. These last years, upstream researches have
resulted in first demonstrators showing the feasibility of such systems \cite{Hong2014CM}. Therefore,
full duplex technology offers a promising potential for 5G. As a result, physical layer
security for full duplex systems is a promising research area that has attracted much
attention recently. The research on full-duplex physical layer security transmission
can be mainly classified into the following four categorizations.

\subsection{Full Duplex Receiver}
W. Li \textit{et al.} first study
a single-antenna transmitter, a two-antenna full duplex receiver,
and a single-antenna eavesdropper wiretap channel \cite{Li2012CLetter}, where
the full duplex receiver use one antenna to receive
the signal and another antenna to send AN to the eavesdropper.
The perfect self-interference cancellation (SIC) is assumed at the receiver.
The closed-form expression of the secrecy outage probability for the proposed transmission scheme
is derived. G. Zheng \textit{et al.} investigate the joint transmit and
receive beamforming design for a single-antenna input, multiple-antenna
output, and multiple-antenna eavesdropper (SIMOME) wiretap channel with imperfect SIC \cite{Zheng2013TSP}.
The full duplex receiver transmits AN to the eavesdropper while receives data from the
transmitter.  For the global perfect CSI assumption, the linear receiver matrix
and the AN generation matrix which maximize the achievable secrecy rate
are jointly designed. It is shown that unlike the half duplex case,
the secrecy rate no longer saturates at high SNR for full
duplex case.  The transmission design for the
statistical CSI of the eavesdropper assumption is also studied in \cite{Zheng2013TSP}.
Moreover, a secure communication scheme for MIMOME wiretap channel with
full duplex receiver and perfect SIC is designed in \cite{Zhou2014SPL}.
L. Li \textit{et al.} further derive a closed-form expression
for the maximal achievable
secure degrees of freedom of the MIMOME wiretap channel with full duplex receiver
under the global perfect CSI and the perfect SIC assumptions \cite{Li2016SPL}.

M. Masood \textit{et al.} study a MIMOME wiretap channel with multiple
full duplex receivers and multiple eavesdroppers
under the global perfect CSI and the imperfect SIC assumptions \cite{Masood2017CLetter}.
Both the transmitter and the full duplex receivers will send AN to degrade
the channels of the eavesdroppers. In this case, the precoding matrix and
the AN generation matrix are optimized jointly to maximize the
achievable secrecy rate. In the mean time, B. Akgun \textit{et al.} consider
a similar multiple-input single-output multiple-antenna eavesdropper (MISOME) wiretap channel, where the transmitter employs a zero-forcing
beamforming (ZFBF) to eliminate the multiuser interference \cite{Akgun2017TCOM}. Then, the total transmit
power is minimized for both instantaneous and statistical CSI of the eavesdroppers cases
subject to the individual secrecy rate constraint for each user.
L. Chen \textit{et al.} investigate single antenna multi-carrier wiretap channels
with full duplex receivers \cite{Chen2017TSP}, where the power allocation among the subcarriers
is designed to maximize the secrecy rate under both the total power and the
legitimate links sum rate constraints. The secure communication in a single-input single-output
multiple-antenna eavesdropper (SISOME) wireless ad hoc network is analyzed in \cite{Zheng2017TWC}, where a hybrid full/half duplex receiver deployment
strategy is employed. The fractions of full duplex receivers which optimize the secure link number,
the network-wide secrecy throughout, and the network-wide secrecy energy efficiency are derived.

For a decentralized heterogeneous network which includes a half duplex receiver tier and a full duplex receiver tier,
T.-X. Zheng \textit{et al.} derive the secrecy outage probability of a typical full duplex receiver
based on the stochastic geometry framework \cite{Zheng2017TWC_2}. In addition, the deployment of the full duplex receivers
is optimized for the network-wide secrecy throughput maximization.
T. Zhang \textit{et al.} design the secrecy communication schemes for a cognitive wiretap channel
with a multiple antenna full duplex secondary receiver \cite{Zhang2017TCOM}.
A brief summary of above work is given in Table \ref{table:FDR}.

\begin{table}
\centering
  \renewcommand{\multirowsetup}{\centering}
 \captionstyle{center}
  {
\caption{Secure Communications with Full Duplex Receivers}
\vspace{0.2cm}
\label{table:FDR}
\begin{lrbox}{\tablebox}
\begin{tabular}{|c|c|c|c|}
\hline
  Paper   & System Model & CSI  & Objective   \\ \hline
 \multirow{2}{*} {W. Li \textit{et al.} \cite{Li2012CLetter}} & A single-antenna transmitter, a two-antenna full duplex receiver & Perfect CSI of the desired user & Analyze  secrecy  \\
 &  A single-antenna eavesdropper, perfect SIC   & Know the noise of eavesdropper & outage probability \\ \hline
  \multirow{2}{*} {G. Zheng \textit{et al.} \cite{Zheng2013TSP}} & A single-antenna transmitter, a multiple-antenna
receiver & Global perfect CSI  &  Joint transmit and \\
 & A multiple-antenna eavesdropper, imperfect SIC   & Statistical CSI of the eavesdropper & receive beamforming design \\ \hline
   \multirow{2}{*} {Y. Zhou \textit{et al.} \cite{Zhou2014SPL}} &    \multirow{2}{*} {MIMOME wiretap channel, perfect SIC} & Perfect CSI of the desired user   &
    Design a secure  \\
 &   & Statistical CSI of the eavesdropper & communication scheme \\ \hline
    \multirow{2}{*} {L. Li \textit{et al.} \cite{Li2016SPL}} &    \multirow{2}{*} {MIMOME wiretap channel, perfect SIC} &  \multirow{2}{*} {Global perfect CSI}  &
     \multirow{2}{*} {Derive secure degrees of freedom} \\
 &   &  & \\ \hline
 \multirow{2}{*} {M. Masood \textit{et al.} \cite{Masood2017CLetter}} & MIMOME, multiple receivers &  \multirow{2}{*} {Global perfect CSI}  &
     \multirow{2}{*} {Design precoding and AN matrices} \\
 & Multiple eavesdroppers, imperfect SIC  &  & \\ \hline
  \multirow{2}{*} {B. Akgun \textit{et al.} \cite{Akgun2017TCOM}} & MISOME, multiple receivers &  Global perfect CSI  &
     \multirow{2}{*} {Minimize total transmit power} \\
 & Multiple eavesdroppers, imperfect SIC  &  Statistical CSI of the eavesdropper &  \\ \hline
   \multirow{2}{*} {L. Chen \textit{et al.} \cite{Chen2017TSP}} & Single antenna multi-carrier  &     \multirow{2}{*} {Global channel amplitudes}  &
    Design power allocation  \\
 & Wiretap channels, imperfect SIC  &   & among the subcarriers \\ \hline
    \multirow{2}{*} {T.-X. Zheng \textit{et al.} \cite{Zheng2017TWC}} & SISOME wireless ad hoc network  &     \multirow{2}{*} {Global perfect CSI}  &
  Optimize full duplex   \\
 &  Imperfect SIC  &   & receivers deployment \\ \hline
    \multirow{2}{*} {T.-X. Zheng \textit{et al.} \cite{Zheng2017TWC_2}} & Two-tie heterogeneous network  &    Perfect CSI of the desired users    &
 Analyze secrecy    \\
 &  Imperfect SIC  &  Statistical CSI of the eavesdroppers  & outage probability \\ \hline
     \multirow{2}{*} {T. Zhang  \textit{et al.} \cite{Zhang2017TCOM}} & A cognitive wiretap channel
with a multiple antenna  &    Perfect CSI of the desired users    &
 Analyze secrecy    \\
& full duplex secondary receiver, perfect SIC  &     No CSI of the eavesdroppers  & outage probability \\ \hline
\end{tabular}
\end{lrbox}
\scalebox{0.8}{\usebox{\tablebox}}
}
\end{table}

\subsection{Full Duplex Transmitter and Receiver}
O. Cepheli \textit{et al.} investigate the bidirectional secure communication where two multiple-antenna
full duplex nodes communicate with each other in presence of a multiple antenna eavesdropper \cite{Cepheli2014CL}.
The global perfect CSI and imperfect SIC assumptions are adopted.
The beamforming vectors are designed to minimize the total transmit power subject to
the secrecy and the QoS constraints.  Assuming the global perfect CSI and perfect SIC,
Y. Wan \textit{et al.} maximize the secrecy sum rate of bidirectional full duplex communication systems in presence of a single-antenna
eavesdropper under the sum transmit power constraint \cite{Wan2015CL}. A null space based suboptimal design
is also proposed to reduce the computational complexity.
 Q. Li \textit{et al.} extend this design to the imperfect SIC and the imperfect CSI of the eavesdropper case \cite{Li2017TSP}.
 A brief summary of above work is given in Table \ref{table:FTR}.

\begin{table}
\centering
  \renewcommand{\multirowsetup}{\centering}
 \captionstyle{center}
  {
\caption{Bidirectional Secure Communications with Full Duplex Transmitters and Receivers}
\vspace{0.2cm}
\label{table:FTR}
\begin{lrbox}{\tablebox}
\begin{tabular}{|c|c|c|c|}
\hline
  Paper   & System Model & CSI  & Objective   \\ \hline
 \multirow{2}{*} {O. Cepheli \textit{et al.} \cite{Cepheli2014CL}} & Multiple-antenna nodes & \multirow{2}{*} {Global perfect CSI} &  Minimize the total transmit power  \\
 & Transmit and receive beamforming, imperfect SIC & & under rate constraints  \\ \hline
  \multirow{2}{*} {Y. Wan \textit{et al.} \cite{Wan2015CL}} & Multiple-antenna transmitter and receiver & \multirow{2}{*} {Global perfect CSI} & Maximize the secrecy sum rate\\
 & Single-antenna eavesdropper, perfect SIC & & under the transmit power constraints  \\ \hline
   \multirow{2}{*} {Q. Li \textit{et al.} \cite{Li2017TSP}} & Multiple-antenna transmitter and receiver & Perfect CSI of the eavesdropper &  Maximize the secrecy sum rate \\
 & Single-antenna eavesdropper, imperfect SIC & Imperfect CSI of the eavesdropper & under the transmit power constraints \\ \hline
\end{tabular}
\end{lrbox}
\scalebox{0.9}{\usebox{\tablebox}}
}
\end{table}

\subsection{Full Duplex Base Station}
\begin{figure*}[!ht]
\centering
\includegraphics[width=0.7\textwidth]{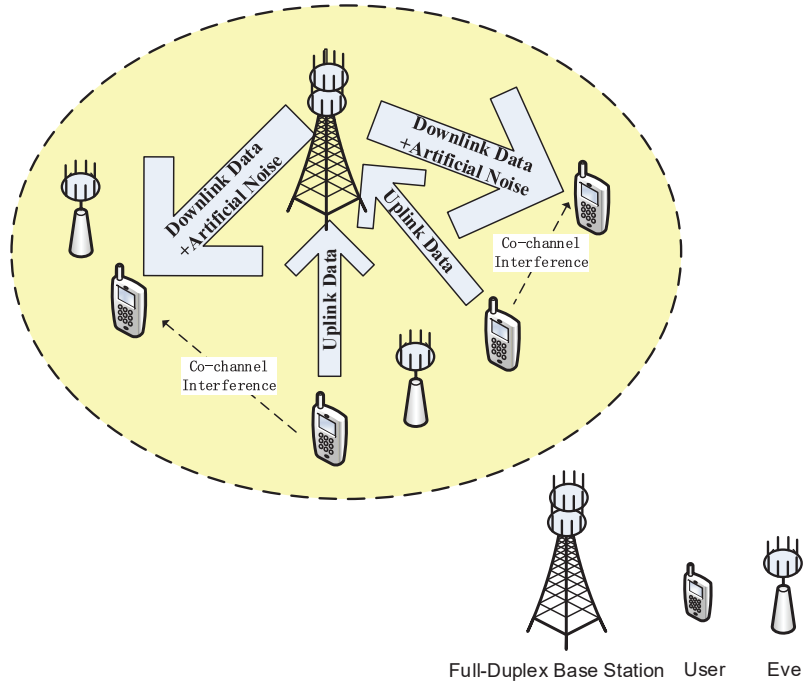}
\caption {\space\space Secure communication for the full-duplex base station network.}
\label{Full_Duplex}
\end{figure*}

\begin{figure*}[!ht]
\centering
\includegraphics[width=0.7\textwidth]{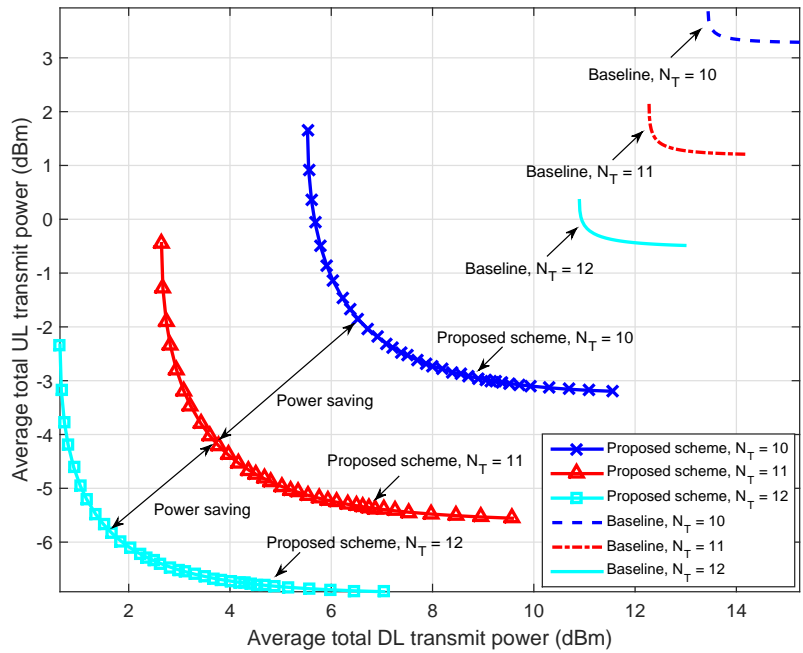}
\caption {\space\space Average trade off between the downlink and uplink total transmit powers.  Experimental results extracted
from \cite{Sun2016TWC}. }
\label{fig_tradeoff_err5}
\end{figure*}

Considering a multiple-antenna full duplex base station which communicates with
a single-antenna transmitter and a single-antenna receiver simultaneously with single-antenna eavesdropper, F. Zhu \textit{et al.}
investigate the joint precoding and AN generation design at the base station with global perfect CSI and perfect SIC
to guarantee both the uplink and downlink transmission security \cite{Zhu2014TSP}.
This work is further extended to the imperfect SIC case \cite{FZhu2016TWC}.
Y. Sun \textit{et al.} study a more general system where a multiple-antenna full duplex base station receives information from
multiple single-antenna uplink users and transmit information to multiple single-antenna
downlink users simultaneously in the presence
of multiple potential eavesdroppers \cite{Sun2016TWC}, as shown in Fig. \ref{Full_Duplex}.
It is assumed that only imperfect CSIs of eavesdroppers are available at the
base station. A robust resource allocation scheme is designed to minimize
a total of uplink and downlink transmit power subject to the uplink and the downlink
rate and security rate constraints. As illustrated in Fig. \ref{fig_tradeoff_err5},
the proposed design achieves significantly higher power efficiency comparing to the baseline ZFBF scheme.
Y. Wang \textit{et al.} investigate the secure transmission for simultaneous wireless
information and power transfer full duplex base station systems \cite{Wang2016SPL}.
A brief summary of above work is given in Table \ref{table:FBS}.

\begin{table}
\centering
  \renewcommand{\multirowsetup}{\centering}
 \captionstyle{center}
  {
\caption{Secure Communications with Full Duplex Base Station}
\vspace{0.2cm}
\label{table:FBS}
\begin{lrbox}{\tablebox}
\begin{tabular}{|c|c|c|c|}
\hline
  Paper   & System Model & CSI  & Objective   \\ \hline
 \multirow{2}{*} {F. Zhu \textit{et al.} \cite{Zhu2014TSP}} & A single-antenna transmitter, a single-antenna receiver & \multirow{2}{*} {Global perfect CSI} & Maximize the secrecy transmit rate under \\
 &  A single-antenna eavesdropper, perfect SIC   & & the secrecy receive rate constraint  \\ \hline
  \multirow{2}{*} {F. Zhu \textit{et al.} \cite{FZhu2016TWC}} &  A single-antenna transmitter, a single-antenna receiver & \multirow{2}{*} {Global perfect CSI} & Minimize the power at base station \\
 & Multiple single-antenna eavesdroppers, imperfect SIC   & & under different  SINR constraints  \\ \hline
   \multirow{2}{*} {Y. Sun \textit{et al.} \cite{Sun2016TWC}} &  Multiple single-antenna transmitters and receivers & Perfect CSI of the legitimate links  &  Minimize the total power \\
 & Multiple multiple-antenna eavesdroppers, imperfect SIC   & Imperfect CSI of eavesdroppers & under secrecy rate constraints  \\ \hline
 \multirow{2}{*} {Y. Wang \textit{et al.} \cite{Wang2016SPL}} &  A single-antenna transmitter, a single-antenna receiver & \multirow{2}{*} {Global perfect CSI}   &  Maximize  the secrecy rate under  \\
 & a single-antenna eavesdropper, imperfect SIC   &  &  power and harvested energy constraints  \\ \hline
\end{tabular}
\end{lrbox}
\scalebox{0.8}{\usebox{\tablebox}}
}
\end{table}

\subsection{Full Duplex Eavesdropper}
Standing at the point of the eavesdropper, A. Mukherjee \textit{et al.} consider a multiple antenna full duplex
active eavesdropper which simultaneously eavesdrops and attacks the legitimate MIMO communication link \cite{Mukherjee2011}.
It is assumed that the eavesdropper has the perfect knowledge of the channels among all nodes and the imperfect
estimation of the self interference channel. Then, the jamming signals
which minimize the secrecy rate are designed based on the Karush--Kuhn--Tucker (KKT) analysis.
X. Tang \textit{et al.} formulate the active eavesdropper problem into a hierarchical game theory problem
where the eavesdropper and the legitimate user behave as a leader and a follower \cite{Tang2017TCom}.
Then, the optimal transmission strategies at both the eavesdropper and the legitimate user's side
are designed. Under the CSI uncertainty condition, M. R. Adedi \textit{et al.} design robust transmission schemes to maximize the secrecy
rate with the multiple antenna full duplex active eavesdropper and the multiple antenna full duplex receiver \cite{Abedi2017TWC}.
A brief summary of above work is given in Table \ref{table:FAE}.

\begin{table}
\centering
  \renewcommand{\multirowsetup}{\centering}
 \captionstyle{center}
  {
\caption{Secure Communications with Full Duplex Active Eavesdropper}
\vspace{0.2cm}
\label{table:FAE}
\begin{lrbox}{\tablebox}
\begin{tabular}{|c|c|c|c|}
\hline
  Paper   & System Model & CSI  & Objective   \\ \hline
 \multirow{2}{*} {A. Mukherjee \textit{et al.} \cite{Mukherjee2011}} & Multiple-antenna nodes & \multirow{2}{*} {Global perfect CSI} &  Minimize the secrecy rate \\
 & Imperfect SIC & &  under a maximum power constraint \\ \hline
  \multirow{2}{*} {X. Tang \textit{et al.} \cite{Tang2017TCom}} & Single-antenna nodes & Perfect CSI of the eavesdropper  & Design optimal strategies within   \\
 & Imperfect SIC & Statistical CSI of the eavesdropper & a hierarchical game theory framework \\ \hline
   \multirow{2}{*} {M. R. Adedi \textit{et al.} \cite{Abedi2017TWC}} & Multiple-antenna nodes & Imperfect CSI between the  & Maximize the secrecy rate    \\
 & Imperfect SIC & eavesdropper and other nodes  & under the transmit power constraint \\ \hline
\end{tabular}
\end{lrbox}
\scalebox{0.9}{\usebox{\tablebox}}
}
\end{table}

There are some other works investigating full-duplex relay secure communications \cite{Lee2015CL,Parsaeefard2015IFS,Chen2015TFS,Li2016JSTSP}.

\section{Other Important Research Work}
Layered approaches have been traditionally
applied for wireless cooperative networks, where each layer in the protocol stack
is designed and operated independently. The motivation for such layered approaches is to
exploit the advantage of the modularity in system design since the system dynamics caused
by the interactions among the protocols at the different layers could be fairly complex.
However, careful exploitation of some cross-layer protocol interactions can lead to a more
efficient performance of the transmission protocol stack and hence better application level
protocol performance in various wireless networking scenarios. This is particularly true for realizing
network security since by exploiting the security capacity and signal processing technologies at
the physical layer and the authentication and watermarking strategies at the application layer,
the available network resources can be utilized more efficiently. A joint physical-application
layer secure transmission scheme which includes physical layer channel coding and application layer authentication and watermarking
is  proposed for multimedia communication systems \cite{Zhou2014CM}. Simulations show that
the joint scheme improves the verification probability for both static and dynamic networks.
On the other hand, to limit the information leakage at
bit-level, the physical layer security technique is exploited to guarantee the
secure video transmission \cite{Hussain2016MTA}. Moreover, physical layer security aware routing for multi-hop ad hoc
networks  are investigated \cite{Ghaderi2015TMC,Yao2016TCom,Xu2016CN}. Other researchers study the physical layer authentication \cite{Borle2015TIFS,Hou2016TCom}.

In addition to theoretical studies, it is also necessary to investigate the
practical test bed design for physical layer security. This is important
to evaluate the usefulness of the physical layer based security schemes and evaluate their performance
in a practical transmission environment. Both WiFi and
LTE test beds are built to examine the secrecy schemes including secrecy coding,  secret key generation,
and artificial noise and beamforming \cite{Phylaws_Wifi,Phylaws_LTE}. It is revealed that in practical transmission environment,
AN is still an effective approach to degrade the performance of
the eavesdropper even when the desired user and the eavesdropper are very close to each other.
Also, it is shown that the channel realizations have a significant impact on
the secrecy coding performance. Other practical prototypes involve more on
secrecy key generation, including in ultra wideband systems \cite{Wilson2007TIFS}, IEEE 802.11 systems \cite{Wei2013TMC,Zhang2016IA},
FM/TV systems \cite{Mathur2011}, etc.

\section{Future Technical Challenges}
In this section, a number of technical challenges for physical layer security in 5G and beyond are
discussed.

\subsection{Physical Layer Security Coding}
As indicated in Table \ref{table:LPDC}, in terms of mutual information criterion,
current LDPC code designs can only achieve the weak secrecy for the special BEC model.
For strong secrecy, it further requires the main channel to be noiseless.
How to design LDPC codes which can achieve the weak/stong secrecy for a more general
channel such as the Gaussian wiretap channel is still a challenge problem.
In terms of BER criterion, the LDPC codes for MIMO and massive MIMO systems can
be investigated.

As indicated in Table \ref{table:Polar}, most of current polar code designs require the
perfect channel knowledge of the eavesdropper at the transmitter to achieve the weak/strong secrecy.
How to extend the polar code designs to a more reasonable case where only channel
distribution knowledge of the eavesdropper is available at the transmitter is an
important research issue. This point also applies for lattice code designs.

\subsection{Physical Layer Security in Massive MIMO Systems}
The transmission designs in \cite{Wu2016TIT} are more effective
to combat the pilot contamination attack for strong correlation channels.
The defense strategy for i.i.d. fading channels in \cite{Basciftci2017} requires
the length of the pilot signal scales with the number of transmit antennas,
which will significantly reduce the transmission efficiency for
massive MIMO systems. The defense strategy in \cite{Im2015TWC} is
used for secret key agreement. Therefore, for massive MIMO transmission with active eavesdropper, the existing approaches are
preliminary and far-from realizing secure communications theoretically and practically for
massive MIMO systems. Several important issues need to be clarified for massive MIMO systems under pilot contamination attack: 1) What is the
fundamental limit of secure communication? 2) Are there any unified transmission schemes which are effective
for general massive MIMO channels?  3) How to design the secure transmission schemes in practical communication
systems?

Since massive MIMO is the key technology for 5G wireless networks,
another important research point for  physical layer security in massive MIMO systems
is the secure transmission scheme designs for the joint implementation of massive MIMO and other important 5G technologies,
e.g., the secure mmWave massive MIMO communications and the secure massive communications in heterogeneous networks, etc.

\subsection{Physical Layer Security for mmWave Communications}
Normally, highly directional beamforming is used in mmWave communications to combat the path loss.
This directional beamforming significantly depends on the accuracy of the CSI of the desired users.
Similar as the massive MIMO case, for TDD communication systems, if an active eavesdropper jeopardizes the channel estimation phase
for the desired user, this may result in a significant secure threat. However, little research has been available
on how to defend the active eavesdropper in mmWave communications.

For a lower implementation cost, hybrid digital and analog precoding
is often used for mmWave communications to reduce the number of RF chains. How to design secure transmission schemes based on
this hybrid structure for both point-to-point and network mmWave MIMO communication systems can be studied.
Secure mmWave vehicle to vehicle communications is another important research point.

\subsection{Physical Layer Security in Heterogeneous Networks}
Currently, most works for physical layer security in heterogeneous networks \cite{Wu2016CLett,Tolossa2017IOT,Xu2016CLett,Wang2016TCOM,Wang2017TFS,Wu2016Globecom} focus on
analyzing the secrecy performance of the networks.  Since multiple tiers are available
to multiple users, another possible research strategy is to investigate how to properly
schedule these users access to different network ties in order to better safeguard the
multi-tier communications. Based on this, precoder designs can be studied to further
improve the secrecy spectrum efficiency. Moreover, the interference generated by multiple
tiers may be properly exploited to degrade the performance of the eavesdropper.

\subsection{Physical Layer Security of NOMA}
The security risk is a particular concern for the multiple access techniques such as NOMA since
the user is allowed to decode the transmit information for other users. As a complement
of current encryption techniques, physical
layer security technology is a good candidate for improving the communication security for
NOMA systems.

Currently, there are only some very initial research results for physical layer security of NOMA \cite{Zhang2016CLetter,Liu2017TWC,Ding2017TCom},
primary focusing on analyzing the secrecy performance of NOMA systems or providing the power allocation policy
for an ideal simple NOMA model. More research work are required to design effective and efficient secure communication
schemes for practical NOMA systems.

\subsection{Physical Layer Security for Full Duplex Technology}
If all the nodes in the systems are capable of full duplex communication ability, how to
design the effective secure communication scheme remains unknown. In particular,
the full duplex transmitter and receiver can generate AN to degrade the eavesdropper,
while the full duplex eavesdropper can also generate AN to interfere
the transmitter and the receiver. This may be formulated into
a hierarchical game framework where some game theory methods can
be exploited to solve the problem.

Another point is that for full duplex base station scenarios, current research assume
both transmitter and receiver are equipped with single-antenna. The extension to multiple-antenna
case can be investigated.

\subsection{Physical Layer Security for Other 5G Scenarios and Beyond}
Physical layer security technique have many other applications in 5G communications and beyond. For example,
traditional security key mechanisms are mainly based on
the distribution of shared keys. Due to the mobility and scalability of 5G wireless networks \cite{5G-PPP}, this task
is nontrivial. The physical layer secret-key generation  was first investigated in \cite{Maurer1993TIT,Ahlswede1993TIT},
where correlated observations of noisy phenomena can be exploited to generate secret keys by exchanging
information over a public channel.  As one of the few implementable physical layer security techniques,
key generation can be constructed in current wireless devices.
Many prototypes have been reported involving physical layer secret-key generation \cite{Wilson2007TIFS,Wilhelm2013JSAC,Patwari2010TMC,Zenger2016}.
However, thorough study examining the effects of environment conditions and
channel parameters on the physical layer secret-key generation is still missing.
In addition, in the case of key agreement, most studies consider
key generation schemes with passive eavesdroppers,
where the active attacks have been less studied.

The internet of thing (IoT) is the network of physical objects embedded with actuators, radio-frequency identifications,
sensors, software, and connectivity to enable it to interact with manufacturers, operators, and/or other connected devices to reach
common goals. 5G will be a key enabler for the IoT by providing the planform to connect a massive number of machine-type communication (MTC) devices to
internet. MTC devices are usually low data rate requirements, periodic data traffic arrivals, limited hardware and signal processing complexity,
limited storage memory, compact form factors, and significant energy constraints \cite{Liang2013JESTCS}. These aspects have received
relatively limited attention on physical layer security in the literature. For example, a theoretically well-founded and holistic approach
to precisely characterize complexity and energy constraints in physical layer security designs is still missing. Moreover,
a network of massive MTC devices in IoT requires novel fundamental definitions of secrecy metrics for point-to-multipoint systems
and multipoint-to-point systems  with a very large number of downlink receivers and uplink transmitters, respectively.
In addition, the communication channels of MTC devices may have very different propagation characteristics as opposed
to the conventional Rayleigh and Rician multipath channel models for broadband microwave systems \cite{Saad2014TWC}.
How to securely transmit data over these channels remains largely open.

Other interesting work includes the interplay between wireless power transfer and physical layer security \cite{Xing2016TWC,Khandaker2016TFS}.
In particular, it is shown in \cite{Xing2016TWC,Khandaker2016TFS} that the jamming noise which can be used to provide security
may not always be harmful and can be an energy beam as well.
Therefore, it may be a good integration between wireless power transfer and physical layer security design.

\section{Conclusion}
The emerging and development of future wireless technologies such as massive MIMO technology,
millimeter wave communications, machine type communication and Internet of Thing, etc have brought out new
security challenges for 5G networks. To design efficient secure transmission schemes
for 5G wireless communications that exploit propagation properties of radio channels in physical layer
has attracted wide research interests recently. This approach is referred as physical layer security for 5G technologies.
The physical layer security approaches are robust to more and more advanced
passive and active eavesdroppers and are flexible for secret key generation in 5G networks.
With careful management and implementation, physical layer security and conventional encryption techniques can
formulate a well-integrated security solution together that efficiently safeguards the confidential and privacy
 communication data in 5G networks. We wish that the research results in this
special issue and our survey paper will be helpful to the readers to have  a better understanding of the benefits and
opportunities that physical layer security techniques provide for 5G and future wireless networks.



\end{document}